\documentclass[prd,eqsecnum,preprint,tightenlines,nofootinbib,showpacs]{revtex4}
\usepackage{graphicx,amsmath}

\def\bea{\begin{eqnarray}}
\def\eea{\end{eqnarray}}
\def\be{\begin{equation}}
\def\ee{\end{equation}}

\begin{document}

\title{Zero momentum modes in discrete light-cone quantization}

\author{Sofia S. Chabysheva}
\affiliation{Department of Physics \\
Southern Methodist University \\
Dallas, Texas 75275 \\
and \\
Department of Physics \\
University of Minnesota-Duluth \\
Duluth, Minnesota 55812}

\author{John R. Hiller}
\affiliation{Department of Physics \\
University of Minnesota-Duluth \\
Duluth, Minnesota 55812}

\date{\today}

\begin{abstract}
We consider the constrained zero modes found in the application of
discrete light-cone quantization (DLCQ) to the nonperturbative
solution of quantum field theories.  These modes are usually
neglected for simplicity, but we show that their inclusion can
be relatively straightforward, and, what is more, that they are
useful for nonperturbative calculations of field-theoretic
spectra.  In particular, inclusion of zero modes improves the
convergence of the numerical calculation and makes possible
the direct calculation of vacuum expectation values, even
when the zero modes are determined dynamically.  We also
comment on zero-mode contributions not included by DLCQ,
namely zero-mode loops.
\end{abstract}

\maketitle

\section{Introduction}
\label{sec:Introduction}

The technique of discrete light-cone quantization
(DLCQ)~\cite{PauliBrodsky,MaskawaYamawaki,Thorn,DLCQreviews} has been
employed many times for the nonperturbative solution of various
field theories, particularly in two dimensions.  This includes
recent calculations of eigenstates in supersymmetric Yang--Mills
theories~\cite{SuperYangMills} and $\phi^4$ theory~\cite{RozowskyThorn,Varyetal},
as well as the early applications to Yukawa theory~\cite{PauliBrodsky},
$\phi^3$ and $\phi^4$ theories~\cite{VaryHari-phi3,VaryHari-phi4},
QED~\cite{QED1+1}, and QCD~\cite{QCD1+1}.
The method is based on light-front quantization~\cite{Dirac}
and discretization of the single-particle momentum modes.  Those modes 
with zero momentum, zero modes, are not dynamical but are instead constrained
by the spatial average of the Euler--Lagrange equation
for the field~\cite{MaskawaYamawaki,Heinzl,Robertson}.  As such, the zero modes 
are usually neglected, because the constraint equation is considered 
too difficult to solve.

The neglect of zero modes can have various consequences, ranging
from the benign, such as slowed convergence of a numerical calculation,
to the serious, an absence of understanding of vacuum 
effects, particularly symmetry 
breaking~\cite{Heinzl,Robertson,Hornbostel,Pinsky,Grange}.  
Can they instead be included in
some straightforward way?  We argue here that they can.  Although
the constraint equation can be a nonlinear operator equation, the
DLCQ approximation requires only a finite expansion of the solution
in inverse powers of the resolution $K$, where discrete
longitudinal momentum fractions are measured in multiples of $1/K$.
The solution to the constraint equation can then be generated
analytically order by order.
For simplicity, we formulate the discussion in two dimensions, 
but the approach can be immediately extended to modes with zero 
longitudinal momentum in any number of dimensions.

In general, the DLCQ approximation is equivalent to a numerical quadrature 
for the field-theoretic mass eigenvalue problem, where the eigenstate
is expanded in Fock states with momentum wave functions as
coefficients.  The eigenvalue problem can be reduced to coupled
integral equations for these wave functions.
The quadrature points, in terms of momentum fractions, are equally 
spaced by $1/K$.  The zero-mode contributions are contributions
from the numerical approximation where a momentum is zero but
the integrand is nonzero, even though the wave function itself
may be zero.  To neglect such contributions is
an error of order $1/K$, which delays convergence as $K\rightarrow\infty$,
relative to the nominal trapezoidal quadrature error of $1/K^2$.  
Within the DLCQ approximation, such contributions appear as zero-mode 
contributions in the Hamiltonian.  These typically take the
form of effective interactions that include zero-mode exchange
between dynamical modes~\cite{Wivoda,Maeno}.  Higher-order
quadrature schemes can also generate effective interactions,
which can be derived from the numerical approximation
to the coupled integral equations for the wave functions.
In addition, the quadrature scheme associated with DLCQ can
generate contributions that go beyond DLCQ, to include effects
from quantum corrections to the constraint equation; we illustrate
this in the case of $\phi^4$ theory by obtaining a contribution
from a zero-mode loop~\cite{Hellerman,ZeroModeLoop}.

In the remainder of the paper, we develop these ideas more fully.
A general discussion is given in Sec.~\ref{sec:ZeroModes},
followed by three specific applications in Sec.~\ref{sec:Applications}.
The approach is summarized in the final Sec.~\ref{sec:Summary}.

\section{Zero Modes in DLCQ}
\label{sec:ZeroModes}

We consider various two-dimensional theories and the contributions
to their light-front Hamiltonians from modes of zero momentum.
Much of the notation is taken from earlier work on zero modes~\cite{Pinsky}.
As light-front coordinates~\cite{Dirac}, we use 
$x^\pm=(x^0\pm x^1)/\sqrt{2}$.  

Each Lagrangian is of the general form
\be
\mathcal{L}=\partial_+\tilde\phi\partial_-\tilde\phi \mp \frac{\mu^2}{2}\tilde\phi^2
-V(\tilde\phi)+\mathcal{L}_{\rm free}^{\rm other},
\ee
with $\tilde\phi$ the scalar field of interest, 
$V$ a generic interaction term which may include other fields,
and $\mathcal{L}_{\rm free}^{\rm other}$ the free Lagrangian for any other
fields.  The mass term includes the possibility of a plus sign, in order
to consider a $\phi^4$ theory with tree-level symmetry breaking.  
For this Lagrangian, the Hamiltonian density is
\be
\mathcal{H}=\pm\frac{\mu^2}{2}\tilde\phi^2+V(\tilde\phi)
+\mathcal{H}_{\rm free}^{\rm other}.
\ee

We apply DLCQ~\cite{PauliBrodsky,DLCQreviews}
by imposing periodic boundary conditions on $\tilde\phi$ in the box $-L/2<x^-<L/2$.
A constant, zero-momentum mode $\phi_0$ is separated from the other modes, so
that we have $\tilde\phi=\phi+\phi_0$ and
\be
\phi=\sum_{n>0}\frac{1}{\sqrt{4\pi n}}\left[e^{ik_n^+x^-}a_n+e^{-ik_n^+x^-}a_n^\dagger\right],
\ee
with $k_n^+=2\pi n/L$ and $[a_n,a_n^\dagger]=1$.
The zero mode, written $\phi_0=\frac{1}{\sqrt{4\pi}}a_0$, is 
constrained~\cite{MaskawaYamawaki,Heinzl,Robertson} by the Euler--Lagrange equation
\be
(2\partial_+\partial_-\pm\mu^2)\tilde\phi=-V'(\tilde\phi),
\ee
which, after integration over the length of the box, yields
\be \label{eq:constraint}
\mp \mu^2\phi_0=\frac{1}{L}\int_{-L/2}^{L/2}V'(\phi+\phi_0)dx^-.
\ee
This constraint is to be solved, to determine $\phi_0$ and, therefore, $a_0$
in terms of the dynamical modes.  

The DLCQ Hamiltonian operator for evolution in light-front time $x^+$ is
\be
\mathcal{P}^-=\int_{-L/2}^{L/2} dx^- \mathcal{H}
  =\pm\mu^2 \frac{L}{4\pi}\left[\Sigma_2 + \frac12 a_0^2\right]
    +\int_{-L/2}^{L/2} dx^- \left[V(\phi+\phi_0)
            +\mathcal{H}_{\rm free}^{\rm other}\right].
\ee
As in \cite{Pinsky}, we define
\be
\Sigma_n=\frac{1}{n!}\sum_{i_1\cdots i_n\neq0}\frac{\delta_{i_1+\cdots+i_n,0}}
     {\sqrt{|i_1\cdots i_n|}}:a_{i_1}\cdots a_{i_n}: .
\ee
The form of $\Sigma_n$ is made compact by using a negative index to
indicate a creation operator; when $i<0$, we have $a_i=a_{|i|}^\dagger$.
The expression for $\mathcal{P}^-$ can be simplified once the full form
is specified and the constraint equation is invoked.
Following Rozowsky and Thorn~\cite{RozowskyThorn}, 
we work with a rescaled Hamiltonian
\be
h=\frac{2\pi}{\mu^2L}\mathcal{P}^-
\ee

The eigenstates of $h$ are constructed as Fock-state expansions at fixed
light-cone momentum $P^+=2\pi K/L$, where $K$ is an integer that sets
the resolution of the calculation~\cite{PauliBrodsky}.  The momenta 
$k_n^+=2\pi n/L$
of the particles in each Fock state must sum to $P^+$, and the indices $n$
must then sum to $K$.  The light-cone momenta, and the integer indices,
must be positive; this limits the number of particles to a maximum of $K$.

The eigenvalue problem for $h$ yields coupled equations for the wave
functions of the Fock-state expansion.  In this context, the zero-mode 
contributions come from integration end points where the momentum is zero.
This can be tracked by starting from the continuum form and discretizing the
integral equations for the wave functions.  The discretization equivalent
to the DLCQ approximation is a trapezoidal approximation to the integrals.
In terms of the momentum fractions $x=k^+/P^+$, an integral from zero to
one is replaced by a sum over discrete points in the integral, at
$x_n=n/K=k_n^+/P^+$, multiplied by the interval size, $1/K$:
\be
\int_0^1 dx f(x,1-x)=\frac{1}{2K}f(0,1)+\frac{1}{K}\sum_{n=1}^{K-1}f(n/K,1-n/K)
         +\frac{1}{2K}f(1,0)+\mathcal{O}(1/K^2).
\ee
The end-point corrections, which are the zero-mode contributions, are then
of order $1/K$ higher than the bulk of the sum that approximates the
integral.  Thus, the zero-mode contributions are equivalent to the addition 
of effective interactions to the Hamiltonian, interactions that include
explicit powers of $1/K$~\cite{Wivoda}, and have the effect, at a minimum,
of improving numerical convergence by restoring the $1/K$ terms missed in ignoring
the end-point corrections of integrals.

In DLCQ these terms are generated by the zero-mode part of the field
as determined by the constraint equation.  For consistency with the
underlying trapezoidal approximation, these
contributions should be kept to no higher in $1/K$ than the corrections
expected.  The order is measured relative to the non-zero-mode parts,
which are typically proportional to $\Sigma_n=\bar\Sigma_n/K^{n/2}$,
where
\be \label{eq:barSigma}
\bar\Sigma_n=\frac{1}{n!}\sum_{i_1\cdots i_n\neq0}\frac{\delta_{i_1+\cdots+i_n,0}}
     {\sqrt{|x_1\cdots x_n|}}:a_{i_1}\cdots a_{i_n}:
\ee
is written explicitly in terms of momentum fractions $x_i$.  A non-zero-mode
contribution of $\Sigma_n$ would then require zero-mode contributions of
no more than order $1/K^{1+n/2}$.
We illustrate this in the following section for various
interaction models.

\section{Applications}
\label{sec:Applications}

\subsection{Wick--Cutkosky Model}

The simplest nontrivial case is an interaction $V=\lambda\tilde\phi|\chi|^2$
with a complex scalar field $\chi$ of mass $m$.
The spectrum of this theory is unbounded from below~\cite{Baym}; this
is obvious classically because the $\phi$ field can acquire a negative value and
an arbitrarily strong $\chi$ field can then drive $V$ to $-\infty$.  An ordinary
DLCQ calculation that excludes zero modes can detect this
structure, but not without careful extrapolation~\cite{Swenson}.  
Here we include zero modes, and a simple variational calculation is all that
is needed to detect the unbounded behavior.

We use antiperiodic boundary conditions for the discretized
$\chi$ field and thereby avoid its zero modes.  This option
is not available for the $\tilde\phi$ field, since it is coupled
to the necessarily periodic square of the $\chi$ field.
The mode expansion of the $\chi$ field is
\be
\chi=\sum_{n>0}\frac{1}{\sqrt{4\pi n}}
      \left[c_n e^{ik_n^+x^-}+d_n^\dagger e^{-ik_n^+x^-}\right].
\ee
The constraint equation (\ref{eq:constraint}) yields
\be
a_0=-2g\Sigma_2^\chi,
\ee
where $g=\lambda/\mu^2\sqrt{4\pi}$ and
\be
\Sigma_2^\chi=\frac12\sum_{n>0}\frac1n\left[c_n^\dagger c_n+d_n^\dagger d_n\right].
\ee
The Hamiltonian reduces to
\be
h=\frac12\Sigma_2 + \frac{m^2}{\mu^2}\Sigma_2^\chi-g^2\left(\Sigma_2^\chi\right)^2
  +\frac{g}{2}\sum_{klm\neq0} \frac{\delta_{k+l+m,0}}{\sqrt{|klm|}}a_k
                            \left[c_l c_m +d_l d_m\right],
\ee
where the sum in the last term includes both positive and negative indices.
The $g^2\left(\Sigma_2^\chi\right)^2$ term has as its origin the zero-mode
contributions $\phi_0^2$ and $\phi_0|\chi|^2$ to the Hamiltonian density.  
The constraint equation specifies that $\phi_0$ is proportional to the negative 
of the average of $|\chi|^2$, making a net negative-definite contribution to the 
energy density.  

To isolate this negative contribution, consider the expectation value of $h$
for the highest Fock state 
$\left(c_1^\dagger\right)^{K-l}\left(d_1^\dagger\right)^l|0\rangle$
of $K$ $\chi$-particles in any charge sector, 
each with the same momentum fraction $1/K$.  The expectation value is
\be
\langle h\rangle=\frac{m^2}{\mu^2}\frac{K}{2}-g^2\left(\frac{K}{2}\right)^2.
\ee
Since this tends to $-\infty$ as $K\rightarrow\infty$, the spectrum
extends to $-\infty$ in the continuum limit.\footnote{This fate is avoided
in Yukawa theory, where $\chi$ is a fermi field, simply because the
identical $x=1/K$ states cannot be populated.}

\subsection{$\phi^3$ Theory}

In $\phi^3$ theory there are, of course, no fields other than $\tilde\phi$, and
the interaction is $V=\frac{\lambda}{3!}\tilde\phi^3$.  The constraint
equation (\ref{eq:constraint}) becomes
\be
-a_0=\frac{g}{2} a_0^2+g\Sigma_2,
\ee
with $g=\lambda/\mu^2\sqrt{4\pi}$.  On use of this constraint, the
Hamiltonian can be written
\be
h=\frac12\Sigma_2+\frac{g}{2}\Sigma_3-\frac{1}{12}a_0^2-\frac{g}{12}a_0^3.
\ee
The non-zero-mode piece of the Hamiltonian, the first two terms, can be
written as $\frac{1}{2K}\bar\Sigma_2+\frac{g}{2K^{3/2}}\bar\Sigma_3$,
where the $\bar\Sigma_n$ are defined in Eq.~(\ref{eq:barSigma}).
Thus the zero-mode contributions we seek are of order $1/K^{5/2}$ at
most.

The constraint equation is solved to a consistent order by taking
$a_0=v_0+v_1/K$ and finding $v_0$ and $v_1$.  From the constraint
equation we have
\be
-v_0=\frac{g}{2}v_0^2\;\; \mbox{and} \;\; -v_1=v_0v_1+v_1v_0+g\bar\Sigma_2.
\ee
For $v_0$, the two possible solutions are $v_0=0$, the local minimum
in the classical energy density, and $v_0=-2/g$, the local maximum.
Either solution is acceptable as a starting point, but the first 
leads to simpler expressions and is sufficient for our purposes.
The second corresponds to a different choice for the perturbative
vacuum, annihilated by the dynamical $a_n$, because it represents
a different choice for $\phi=\tilde\phi-\phi_0$; however, this change
in the perturbative vacuum
is compensated by different terms in the Hamiltonian, with no
net effect on the spectrum.  With $v_0=0$,
the solution for $v_1$ is then immediately $v_1=-g\bar\Sigma_2$,
and we find for the zero mode
\be
a_0=-\frac{g}{K}\bar\Sigma_2+\mathcal{O}(\frac{1}{K^2}).
\ee

On substitution of this expansion for the zero mode, the Hamiltonian 
becomes
\be
h=\frac{1}{2K}\bar\Sigma_2+\frac{g}{2K^{3/2}}\bar\Sigma_3-\frac{g^2}{12K^2}\bar\Sigma_2^2
\ee
The expectation value of $h$ for the state populated with $K$ particles of
momentum fraction $1/K$ is
\be
\langle h\rangle=\frac{K}{2}-\frac{g^2}{12}K^2.
\ee
This tends to $-\infty$ in the $K\rightarrow\infty$ limit, and, as
known for a cubic theory~\cite{Baym}, the spectrum is unbounded from below.

\subsection{$\phi^4$ Theory}

For $\phi^4$ theory, with its interaction $V=\frac{\lambda}{4!}\tilde\phi^4$,
the contributions of zero modes are more subtle than for the first two 
applications.  Their inclusion will improve numerical convergence and
may provide a means to understand vacuum structure and symmetry breaking.
Previous calculations~\cite{RozowskyThorn,Varyetal} of the spectrum
did not include zero modes, and previous studies of the constraint
equation~\cite{Pinsky} attempted to solve it fully, rather than keeping
$a_0$ only to an order consistent with the DLCQ Hamiltonian.

The constraint equation (\ref{eq:constraint}) in this case is
\be
\mp a_0=\frac{g}{3}a_0^3+2g\Sigma_3+\frac{2g}{3}
          \left(a_0\Sigma_2+\Sigma_2 a_0\right)
  +\frac{g}{3}\sum_{n\neq0}\frac{1}{|n|}a_na_0a_{-n},
\ee
where $g=\lambda/8\pi\mu^2$.  The Hamiltonian is
\bea
h&=&\pm\frac{1}{2K}\bar\Sigma_2+\frac{g}{K^2}\bar\Sigma_4
-\frac{g}{24}a_0^4
+\frac{g}{24K^{3/2}}\sum_{klm\neq0}\frac{\delta_{k+l+m,0}}
                                   {\sqrt{x_{|k|} x_{|l|} x_{|m|}}}
                   (a_k a_l a_0 a_m + a_k a_0 a_l a_m) \nonumber \\
&&+\frac{g}{24K}\sum_{n\neq0}\frac{1}{x_{|n|}}
                                 (a_na_0^2a_{-n}-a_0 a_n a_{-n}a_0).
\eea
From the $K$ dependence of the non-zero-mode pieces of the 
Hamiltonian, we see that we need zero-mode corrections to 
order $1/K^3$ and thus expand $a_0$ to order $1/K^{3/2}$.

To solve the constraint equation to this order, we write 
$a_0=v_0+v_1/K^{1/2}+v_2/K+v_3/K^{3/2}$
and solve for the $v_i$.  Two of the three possible 
solutions are available only for the wrong-sign
mass term.  They have $v_0=\pm\sqrt{3/g}$ and 
correspond to the local minima in the classical energy density.  
The third solution,
$v_0=0$, yields the local maximum as well as simpler expressions; 
it is also the only solution in the case of the correct-sign mass term.  
For simplicity, we choose to
work with this third solution, for which $v_1$ and $v_2$ are also zero and 
\be \label{eq:phi4soln}
a_0=\mp\frac{2g}{K^{3/2}}\bar\Sigma_3=\mp 2g\Sigma_3.
\ee
Here the upper (lower) sign corresponds to the upper (lower) sign
in the mass term of the Hamiltonian.

In the Hamiltonian, this solution to the constraint equation, combined with the keeping
of terms to consistent order, yields
\be \label{eq:newphi4hamiltonian}
h=\pm\frac{1}{2K}\bar\Sigma_2+\frac{g}{K^2}\bar\Sigma_4
\mp\frac{g^2}{12K^3}\sum_{klm\neq0}\frac{\delta_{k+l+m,0}}{\sqrt{x_{|k|} x_{|l|} x_{|m|}}}
                   (a_k a_l \bar\Sigma_3 a_m + a_k \bar\Sigma_3 a_l a_m)+\mathcal{O}(1/K^4).
\ee
The first term is the (rescaled) mass term, and the second is the ordinary
interaction term.
The third term provides the zero-mode corrections to the mass eigenvalue problem
at an order in $1/K$ that is consistent with the DLCQ approximation.  The sign
of the term is particularly significant.  In the case of a positive mass term,
this correction is negative, which will allow the spectrum to extend below
zero and provide a nontrivial vacuum state and symmetry breaking at larger 
couplings.  For the case of a negative mass term, where the symmetry breaking
effect is built in, the zero-mode term is positive and can play a role in
restoring the broken symmetry and the trivial vacuum at larger couplings.

It is known~\cite{Hellerman,ZeroModeLoop} that for $\phi^4$ theory there is an
additional $1/K$ correction from a zero-mode loop.  Such loops do not
enter into DLCQ but were found in discretization of perturbative
contributions to forward scattering~\cite{Hellerman}.  
They can also be found from a
nonperturbative perspective by considering the coupled equations
for the $n$-particle Fock-state wave function $\psi_n(x_1,\ldots,x_n)$,
which are
\bea  \label{eq:coupledeqns}
\lefteqn{\sum_{i=1}^n\frac{1}{x_i}\psi_n+\frac{g/3}{\sqrt{n(n-1)}}
\sum_{i\neq j\neq k}\frac{\psi_{n-2}(x_1,\ldots,x_i+x_j+x_k,\ldots,x_n)}
                {\sqrt{x_ix_jx_k(x_i+x_j+x_k)}} } &&    \\
   && + \frac{g}{3}\sqrt{(n+1)(n+2)}\sum_i
       \int \frac{dx'_1 dx'_2
        \psi_{n+2}(x_1,\ldots,x'_1,\ldots,x'_2,\ldots,x_i-x'_1-x'_2,\ldots,x_n)}
                          {\sqrt{x'_1 x'_2 x_i (x_i-x'_1-x'_2)}}
                 \nonumber \\
   && + \frac{g}{2} \sum_{i\neq j}
     \int \frac{dx' \psi_n(x_1,\ldots,x',\ldots,x_i+x_j-x',\ldots,x_n)}
          {\sqrt{x_i x_j x' (x_i+x_j-x')}}
                =(M^2/\mu^2)\psi_n .
     \nonumber
\eea
One of the contributions to $\psi_{n+2}(y_i)$ is
\be
\frac{g/3}{M^2/\mu^2-\sum_i^{n+2}\frac{1}{y_i}}\frac{1}{\sqrt{(n+2)(n+1)}}
   \sum_{i\neq j \neq k} \frac{\psi_n(y_1,\ldots,y_i+y_j+y_k,\ldots,y_n)}
                {\sqrt{y_i y_j y_k (y_i+y_j+y_k)}} ,
\ee
and substitution into Eq.~(\ref{eq:coupledeqns}) yields a term of the form
\bea
\lefteqn{\left(\frac{g}{3}\right)^2\sum_i 
    \int \frac{dx'_1 dx'_2}{\sqrt{x'_1 x'_2 x_i (x_i-x'_1-x'_2)}}
    \frac{1}{M^2/\mu^2-\sum_{j\neq i}\frac{1}{x_j}
                     -\frac{1}{x'_1}-\frac{1}{x'_2}-\frac{1}{x_i-x'_1-x'_2}}}&& \nonumber \\
&&\times\sum_{k \neq i}\frac{\psi_n(x_1,\ldots,x_i-x'_1-x'_2,\ldots,x_k+x'_1+x'_2,\ldots,x_n)}
                           {\sqrt{x'_1 x'_2 x_k (x'_1+x'_2+x_k)}} ,
\eea
which represents the exchange of two bosons, with momentum fractions $x'_1$
and $x'_2$, between the $i$th and $k$th particles of the $n$-body Fock state.
In the corner of the
two-dimensional integral, where both $x'_i$ approach zero and therefore
two zero modes are exchanged, the integrand is singular.  However, this
is an integrable singularity; the net effect is for there to be a
contribution of order $1/K$ from this corner, instead of the nominal
order of $1/K^2$, which DLCQ would neglect.  Where only one $x'_i$
approaches zero, which represents an exchange of one zero mode and
one non-zero mode, we have a $1/K$ contribution that is included by
DLCQ.  This contribution is generated by the new third term in the Hamiltonian
(\ref{eq:newphi4hamiltonian}).  A term to represent the zero-mode
loop contribution can also be added to the Hamiltonian~\cite{Taniguchi}.          

In addition to the zero-mode corrections to the Hamiltonian, 
the zero mode also allows for
a direct calculation of the DLCQ approximation for the vacuum expectation
value.  The eigenstates of $h$ separate into sectors of odd and even particle
number, and in the continuum limit the lowest state of each sector become degenerate.
Let $|o\rangle$
and $|e\rangle$ be the odd and even ground states and form maximally
mixed states $|\pm\rangle=(|o\rangle\pm|e\rangle)/\sqrt{2}$.  The
expectation value for the field is then
\be
v_\pm=\frac{\langle\pm|\tilde\phi|\pm\rangle}{\langle\pm|\pm\rangle}\equiv\pm v.
\ee
Since $\tilde\phi$ changes particle number by one, only the cross
terms contribute to the numerator, and since the states have the
same momentum, only the zero mode of the field can contribute.
Thus, we have
\be
v=\frac{\langle e|\phi_0|o\rangle+\langle o|\phi_0|e\rangle}
                     {\langle e|e\rangle+\langle o|o\rangle}.
\ee
On substitution of the solution (\ref{eq:phi4soln}) to the constraint equation, 
this becomes
\be
v=\mp\frac{g}{\sqrt{\pi}}\frac{\langle e|\Sigma_3|o\rangle+\langle o|\Sigma_3|e\rangle}
                     {\langle e|e\rangle+\langle o|o\rangle}.
\ee
So, one can calculate $v$ if the mass eigenvalue problem is solved to find the
lowest odd and even states, and one can study the continuum limit as
the resolution $K\rightarrow\infty$.

\section{Summary}
\label{sec:Summary}

Although zero modes are traditionally neglected in DLCQ calculations, they can
actually be taken into account without much additional effort.  Each mode
satisfies a constraint equation that is the (light-front) spatial average
of an Euler--Lagrange equation.  The constraint connects the zero mode to
the dynamical degrees of freedom and can be solved either explicitly or
in terms of an expansion in inverse powers of the DLCQ resolution $K$.
The latter expansion is truncated at the order consistent with the DLCQ
approximation.

From the zero-mode solution, one can construct contributions to the
Hamiltonian which will, at a minimum, repair the convergence of the
DLCQ calculation.  The DLCQ approximation is essentially equivalent to a 
trapezoidal approximation to integral equations for Fock-state wave functions,
with quadrature points spaced equally by $1/K$.  The error for such an
integral is of order $1/K^2$, but when the zero modes are neglected
the error is $1/K$ and convergence as $K\rightarrow\infty$ is slowed.  The
zero-mode contributions to the Hamiltonian repair this and restore
the $1/K^2$ behavior of the integration errors.  For some theories,
such as $\phi^4$ theory, there are additional corrections beyond DLCQ
but still of order $1/K$, such as zero-mode loops, that correspond
to order-$\hbar$ quantum corrections to the zero-mode constraint
equation~\cite{Hellerman,ZeroModeLoop}.

Inclusion of zero modes also makes possible the direct calculation
of vacuum expectation values.  Explicit symmetry breaking will yield
a c-number contribution to the zero mode which trivially has a vacuum
expectation value, but this is not the only way to have a nonzero
value.  The zero mode will, in general, have dynamical contributions.
For these, the expectation value will be zero with respect to the
perturbative vacuum.  
However, the spectrum of the theory can be such that the perturbative vacuum
is not the eigenstate of lowest energy.  Instead, some nontrivial 
eigenstate has a lower energy and, with respect to this state,
the dynamical contributions to the zero mode can have a
nonzero expectation value.

\acknowledgments
This work was supported in part by the Department of Energy
through Contract No.\ DE-FG02-98ER41087.


\end{document}